\documentclass[twocolumn,showpacs,aps,prl]{revtex4}
\usepackage{dcolumn} 
\usepackage{graphicx}
\usepackage{subfigure}
\usepackage{epsfig}
\usepackage{epstopdf}
\begin{document}
\title{Dielectric constant boost in  amorphous sesquioxides}

\author{Pietro Delugas,$^{1,2}$ Vincenzo Fiorentini,$^{1,3,4}$ and Alessio Filippetti$^{3}$}

\affiliation{1) NXP Semiconductors,  Leuven, Belgium\\ 2) CNR-IMM, Catania, Italy\\3) CNR-INFM-SLACS, Cagliari, Italy\\ 4)  Dipartimento di  Fisica, Universit\`a di Cagliari, Italy}
\date{\today}
\begin{abstract} 
High-$\kappa$ dielectrics for insulating layers are a current key ingredient of microelectronics.  X$_2$O$_3$ sesquioxide compounds are among the candidates. Here we show  for a typical material of this class, Sc$_2$O$_3$, that the relatively modest dielectric  constant  of its crystalline phase is {\it enhanced} in the   amorphous phase  by over 40 \% (from $\sim$15 to $\sim$22).  This is due to the disorder-induced activation of low frequency cation-related modes which are inactive or inefficient in the crystal, and by the conservation of effective dynamical charges (a measure of atomic polarizability). The analysis employs  density-functional energy-force and perturbation-theory calculations of  the dielectric response of amorphous samples generated by pair-potential molecular dynamics.
\end{abstract} 

\maketitle
High-dielectric constant ("high-$\kappa$") materials are gradually replacing silica \cite{hik} as gate dielectrics in integrated electronics.  Policrystalline layers thereof usually behave badly in electrical terms due to grain-boundary conduction, so that  amorphous layers are seriously studied as well. Whether or not the $\kappa$ of the crystal will be conserved  upon amorphization is not obvious a priori \cite{amorfi}, and should be assessed  during the validation of a candidate dielectric to be used in thin insulating coatings.  

Considerable attention has been devoted recently \cite{expx2o3,lutz} to sesquioxides  X$_2$O$_3$ with the trivalent cation X=Y, Sc, or a rare-earth atom.  Almost all these materials adopt the cubic bixbyite crystal structure. Experiments \cite{expx2o3,lutz} and theory \cite{lutz,bix} largely agree that their static dielectric constant in that phase is in the order of 12 to 15, and of vibrational origin for  about $\sim$60-70\%. The  microscopic rationale for this behavior was given recently by the present authors \cite{lutz,bix}: the infra-red (IR) active vibrational modes are relatively high-frequency and essentially cation-independent motions.   

 As we discussed in detail elsewhere \cite{lalo,rev,dysco,amorfi}, the dielectric constant $\kappa_s$=${\kappa}_{\infty}$+$\kappa_{\rm ion}$ in  high-$\kappa$ materials is generally dominated  by the lattice-vibrational  component ${\kappa}_{\rm ion}$, a factor 2 to 10 larger than  the electronic ${\kappa}_{\infty}$. 
 By its definition \begin{equation}
\kappa^{\alpha\beta}_{\rm ion} = \frac{4\pi e^2}{\Omega}
\sum_{\lambda} 
\frac{z_{\lambda\alpha} z_{\lambda\beta}}{\omega^2_{\lambda}},
\ \ \   
z_{\lambda\beta}=\sum_{i\beta} \frac{Z^*_{i,\alpha\beta}
 \, \xi_{i,\lambda\beta}}{\sqrt{M_i}},
\end{equation}
(with  $z$  the mode charge vector, $\Omega$ the system volume, ${Z}^*_{i,\alpha,\beta}$ the effective or dynamical or Born charge tensor 
and $M_i$ the mass of atom $i$, 
$\xi_{i,\lambda\beta}$ and $\omega$ the  eigenvector and eigenfrequency of mode
$\lambda$ at zero wave-vector, and $\alpha$, $\beta$ cartesian indexes), a large ${\kappa}_{\rm ion}$ is due to  large  effective polarizabilities as measured by dynamical charges, termed "anomalous" if they  exceed nominal ionicity, and to   soft IR  vibrational modes. A question that naturally arises is, will amorphization, i.e. microscopic disorder, deteriorate the dielectric constant  of these oxides ?  In this Letter we show, based on ab initio calculations, that scandia, Sc$_2$O$_3$, has a {\it larger} dielectric constant in its amorphous phase than in its crystal phase. This dielectric constant boost is due to a) hardly any  loss in dynamical  polarizability, and b) disorder-induced IR activation of non-polar low-energy modes related to cation-oxygen combined motions.   Previously \cite{lutz,bix} we found that the  vibrational dielectric behavior in cubic crystalline  sesquioxides is dominated by oxygen motions and essentially independently of the cation. Therefore   the present predictions for scandia are highly likely to apply to all sequioxides.  Further, we point out that this behavior is not unique: recently, we have shown \cite{amorfi} that also RAO$_3$ oxides (R a rare-earth and A a trivalent cation) roughly conserve their dielectric constant in the amorphous phase.

Amorphous Sc$_2$O$_3$ samples were generated by a melt-and-quench   procedure using empirical-potential molecular dynamics as implemented in the  GULP code \cite{gulp}. A periodic 80-atom, cubic, crystalline Sc$_2$O$_3$ sample was molten, and the liquid was equilibrated at T=5000 K for 200 ps to achieve fully randomized  cation positions. It was then quenched down to 500 K by  Nose' constant-T dynamics at a cooling rate 20 K/ps, in 5-ps costant-T intervals separated by abrupt 100 K down-steps.  At  T=500 K, structural properties  (well converged \cite{cool} at this cooling rate)  of the amorphous sample were sampled in a 5 ps run. Tests on the cell size  (80 to 320 atoms) did not indicate appreciable changes in   the microscopic structure. 

After bringing the  system  to 0 K by damped dynamics, the structure and volume were optimized (for the 80-atom sample only) using the ab initio density-functional theory code Espresso \cite{espresso}. Dynamical charges were computed on the final structure by density functional perturbation theory \cite{dfpt}. Phonon frequencies at zero wavevector were obtained from the force constant matrix calculated by finite-difference force calculations (with cartesian displacements of $\pm$0.01 bohr). The technical details of the DFT calculation \cite{espresso} are closely similar to those of  our previous investigation on RAO$_3$ perovskite-derived amorphous samples \cite{amorfi}.  DFT calculations are  indicated hencefort by the acronym GGA (i.e. generalized-gradient approximation \cite{espresso}).

\begin{figure}[ht]
\includegraphics[clip,width=8.5cm]{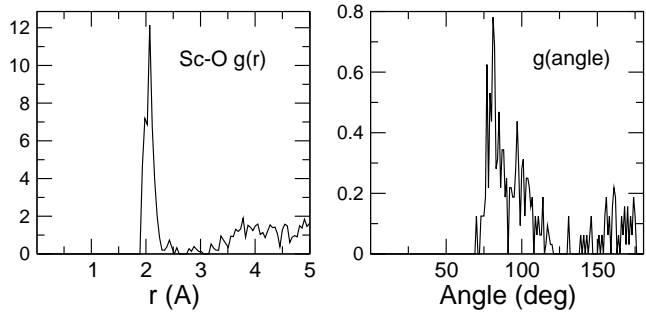}
\caption{\small{Pair correlation functions for the Sc-O pair (left panel) and distribution of O-Sc-O bond angle (right panel) for Sc$_2$O$_3$.}} \label{fig1}
\end{figure}

In bixbyite sesquioxides, the cations are all six-fold coordinated, one out of four in the 
primitive cell has six oxygen neighbors at a single distance, the others have three pairs of neighboring O's at three distinct distances differing by about 0.1 \AA. In amorphous scandia, as shown by Fig. 1, left panel, there is still a quite well defined Sc-O peak. Its width is about twice the crystal value. Correlations at larger distances are washed out by disorder. The same goes for the Sc-O bond angle distribution, which peaks at about 80$^{\circ}$, shows an approximate bimodal structure (at about 80 and 100$^{\circ}$), and is appreciably non-zero all the way from 70$^{\circ}$ to 120$^{\circ}$, indicating a strong orientational disorder or the octahedra making up the structure.

 The dynamical polarizability of vibrating ions  is correctly measured by the effective dynamical charge tensor 
$Z^*_{i,\alpha\beta}$=$\Omega$/e ($\partial P_{\alpha}$/$\partial u_{\beta}$), i.e. the polarization derivative with respect to  atomic displacement \cite{zstar}. A loss of effective polarizability will to decrease the ionic dielectric activity. The  GGA dynamical charges of the amorphous and crystal phases  in Table \ref{tab1} are seen to be essentially the same.  The amorphous-phase charge and dielectric tensors are averaged over  all atoms in a given species and decomposed into  $s$, $p$, and $d$  components \cite{amorfi}; Table \ref{tab1} reports the $s$ component, which contributes in excess of  98\% of the tensor norm, is reported  for the amorphous case. 
 
\begin{table}[htb]
  \centering
 \caption{Average $s$ components of the effective-charge tensor, and of the  electronic, ionic and total static dielectric constant tensors in amorphous and bixbyite-crystalline Sc$_2$O$_3$. The slight difference in oxygen charges is related to differences in the (minor) violations of the dynamical charge neutrality sum rule.}

\begin{tabular}{cccccc}
\hline\hline    
\multicolumn{1}{c}{Sc$_2$O$_3$}&
\multicolumn{1}{c}{$<$$Z_{\rm Sc}^*$$>$}&
\multicolumn{1}{c}{$<$$Z_{\rm O}^*$$>$}&
\multicolumn{1}{c}{$\kappa_{\infty}$}&
\multicolumn{1}{c}{$\kappa_{\rm ionic}$}&
\multicolumn{1}{c}{$\kappa_{s}$}\\
    \hline 
  crystal       & 3.82 &      --2.54 & 4.6 & 10.4 &15 \\
 amorphous     &   3.82 &  --2.53 & 4.5 & 17.5& 22 \\
 \hline\hline   \end{tabular}
    \label{tab1}
\end{table}

The electronic dielectric constant  is almost unaffected by disorder; the  amorphous phase indeed has a crystal-like gap in the electronic density of states, as expected from ionicity and from the anion (cation) character of valence (conduction) states. The static value in the amorphous phase is 40\% larger than the crystal value. The ionic component is larger in the amorphous. To understand why this occurs, we  analyze the frequency-dependent dielectric intensity, namely the dimensionless individual terms in the first equality of Eq.1. Before that, we note in passing that experiments have indeed shown a comparable enhancement: unfortunately,  scandia  recrystallizes during post-growth thermal treatments, so the enhancement effect is destroyed. Recrystallization occurs for most sesquioxides \cite{recry}; whenever it doesn't, an enhancement is indeed observed \cite{recry2}

    \begin{figure}[ht]
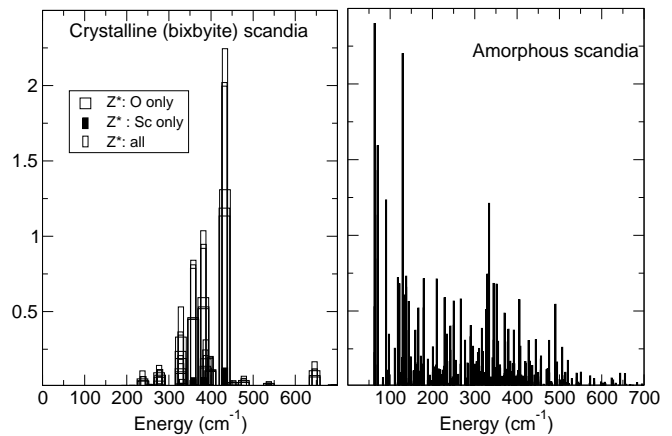

\centering
\includegraphics[clip,width=4.37cm]{./fig2a.eps}
\includegraphics[clip,width=4.18cm]{./fig2b.eps}
\caption{\small{Dielectric intensity for crystalline (bixbyite) scandia (left panel) and  amorphous scandia (right).}}\label{fig2}
\end{figure}

\begin{figure}[ht]   
\centering
\includegraphics[clip,width=7cm]{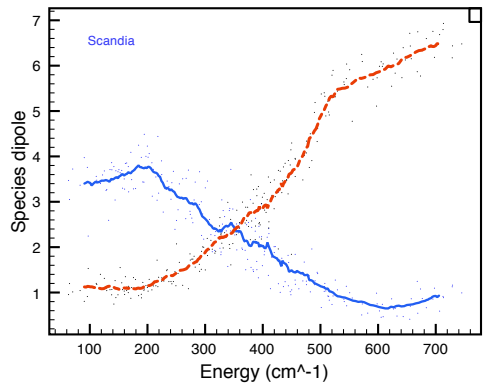}
\caption{\small{Average atom-specific dipoles  in amorphous scandia Sc$_2$O$_3$ (solid line: Sc; dashed: O). Curves are obtained as window averaging (smoothing) of the data points shown as dots.}} \label{fig3}
\end{figure}

 In Fig.\ref{fig2} we compare the dielectric intensity of crystalline bixbyite-structure (right) and amorphous (left) Sc$_2$O$_3$.   Visibly, in Sc$_2$O$_3$ as in other bixbyites the  main IR modes fall around  300-400 cm$^{-1}$. As can be appreciated from the decomposition, these modes are almost entirely oxygen-related; modes involving  cation motions have negligible intensity at low energy, and small in the active IR range (similarly, despite the different intensity ratios, to e.g. Lu$_2$O$_3$ \cite{lutz}). In the amorphous phase, the dielectric enhancement at low frequency is due to IR-inactive (generally Raman) or inefficient crystal modes that acquire IR character (or enhance it) due to disorder: as  Eq.1 shows, the dielectric intensity at low frequency can be dramatically  amplified due to a 1/$\omega^2$ factor provided a non-negligible intensity is present. These new modes are mainly Sc-related: this is evident from 
 Fig.\ref{fig3}, which reports the dipole amplitudes decomposed by atomic species (essentially the "charge vectors" $z_{\lambda}$ defined in Eq.1  obtained summing over one species only). Low frequency modes are seen to be dominated by Sc motions.

In summary,  
we have  shown that the static dielectric response  of amorphous scandia Sc$_2$O$_3$  is enhanced compared  to that of its crystal phase, due to a combination of  polarizability  conservation (effective charges) and of disorder-induced IR activation of inefficient cation-related modes. Whereas  their crystal phase has poor dielectric constant ($\sim$15), we suggest that  amorphous X$_2$O$_3$ materials can typically exhibit intermediate dielectric-constant values ($\sim$22) in the  amorphous phase. 

Work partially supported by  EU (project FUNC@NXP Leuven), MiUR Italy  under PON-Cybersar, Fondazione Banco di Sardegna. Most calculations were done on the SLACS cluster at  CASPUR Rome.

\end{document}